\documentclass[aps,prb,reprint,superscriptaddress,amsfonts,amsmath,amssymb,showpacs,floatfix]{revtex4-1}

\usepackage{graphicx}
\usepackage{multirow}
\usepackage{hyperref}
\newcommand{\mrm}[1]{\mathrm{#1}}

\newcommand{\mbb}[1]{\mathbb{#1}}
\newcommand{\mc}[1]{\mathcal{#1}}
\newcommand{\eref}[1]{(\ref{#1})}
\newcommand{\fref}[1]{Fig.~\ref{#1}}
\newcommand{\tref}[1]{Table~\ref{#1}}
\newcommand{\sref}[1]{Sec.~\ref{#1}}

\newcommand{\ra}{\rangle}
\newcommand{\la}{\langle}
\newcommand{\p}[1]{\phantom{#1}}
\newcommand{\rcite}[1]{Ref.~\onlinecite{#1}} 
\newcommand{\prbtext}[1]{#1}
\newcommand{\pratext}[1]{}
\newcommand{\textcitecomma}[1]{\citeauthor{#1},\cite{#1}}
\newcommand{\citecomma}[1]{,\cite{#1}}
\newcommand{\citestop}[1]{.\cite{#1}}

\newcommand{\Mop}{M_\alpha^{\p{\alpha}\beta}}
\newcommand{\alphabeta}{_\alpha^{\p{\alpha}\beta}}

\begin{document}

\title{Simulation of anyons with tensor network algorithms}

\author{R. N. C. Pfeifer}
\email[]{pfeifer@physics.uq.edu.au}
\author{P. Corboz}
\affiliation{School of Mathematics and Physics, The University of Queensland, St Lucia, QLD 4072, Australia}
\author{O. Buerschaper}
\author{M. Aguado}
\affiliation{Max-Planck-Institut f\"ur Quantenoptik. Hans-Kopfermann-Stra\ss{}e 1, D-85748 Garching, Germany}
\author{M. Troyer}
\affiliation{Theoretische Physik, ETH Zurich, 8093 Zurich, Switzerland}
\author{G. Vidal}
\affiliation{School of Mathematics and Physics, The University of Queensland, St Lucia, QLD 4072, Australia}

\date{\today}

\begin{abstract}
Interacting systems of anyons pose a unique challenge to condensed matter simulations due to their non-trivial exchange statistics. These systems are of great interest as they have the potential for robust universal quantum computation, but numerical tools for studying them are as yet limited. We show how existing tensor network algorithms may be adapted for use with systems of anyons, and demonstrate this process for the 1-D Multi-scale Entanglement Renormalisation Ansatz (MERA). We apply the MERA to infinite chains of interacting Fibonacci anyons, computing their scaling dimensions and local scaling operators. The scaling dimensions obtained are seen to be in agreement with conformal field theory. The techniques developed are applicable to any tensor network algorithm, and the ability to adapt these ans\"atze for use on anyonic systems opens the door for numerical simulation of large systems of free and interacting anyons in one and two dimensions.
\end{abstract}
\pacs{05.30.Pr, 73.43.Lp, 02.70.-c, 03.65.Vf}

\maketitle


\section{Introduction}

The study of anyons offers one of the most exciting challenges in contemporary physics. Anyons are exotic quasiparticles with non-trivial exchange statistics, which makes them difficult to simulate. However, they are of great interest as some species offer the prospect of a highly fault-tolerant form of universal quantum computation\citecomma{kitaev2003,nayak2008} and it has been suggested\prbtext{\cite{xia2004}} that the simplest such species may appear in the fractional quantum Hall state with filling fraction $\nu$ = 12/5\pratext{ \cite{xia2004}}. Despite the current strong interest in the development of practical quantum computing, our ability to study the collective behaviour of systems of anyons remains limited.

The study of interacting systems of anyons using numerical techniques was pioneered by \textcitecomma{feiguin2007} using exact diagonalisation for {1-D} systems of up to 37 anyons, and the Density Matrix Renormalisation Group algorithm (DMRG)\pratext{ }\cite{white1992} for longer chains. Also related is work by \textcitecomma{sierra1997} later extended by \textcitecomma{tatsuaki2000} which applies a variant of DMRG to spin chain models having $SU(2)_k$ symmetry. Some of these models are now known to correspond to $SU(2)_k$ anyon chains\citecomma{trebst2008} and using this mapping these systems may also be studied using the Bethe ansatz\pratext{ }\cite{alcaraz1987} and quantum Monte Carlo\citestop{todo2001}.

However, all of these methods have their limitations. Exact diagonalisation has a computational cost which is exponential in the number of sites, strongly limiting the size of the systems which may be studied. DMRG is capable of studying larger system sizes, but is typically limited to {1-D} or quasi-{1-D} systems (e.g. ladders). Mapping to a spin chain is useful in one dimension but is substantially less practical in two.
There are therefore good reasons to desire
a formalism which will allow the application of other tensor network algorithms 
to systems of anyons. 
Many of these tensor networks, such as 
Projected Entangled Pair States (PEPS)\citecomma{verstraete2004,nishino1998,gu2008,jiang2008,jordan2008}
and the 2-D versions of Tree Tensor Networks (TTN)\pratext{ }\cite{tagliacozzo2009} and of the Multi-scale Entanglement Renormalisation Ansatz (MERA)\pratext{ }\cite{cincio2008,evenbly2009b,evenbly2010}
have been designed specifically to accurately describe two-dimensional systems. 

In one dimension, 
many previously studied systems of interacting anyons display extended critical phases\pratext{ }\prbtext{,}\cite[e.g.][]{feiguin2007,trebst2008}\pratext{,} which are characterised by correlators exhibiting polynomial decay\citestop{difrancesco1997} 
Whereas DMRG favours accurate representation of short range correlators at the expense of long-range accuracy, 
the {1-D} MERA\pratext{ }\cite{vidal2007,vidal2008a} is ideally suited to this situation as its hierarchical structure naturally encodes the renormalisation group flow at the level of operators and wavefunctions\citecomma{vidal2007,vidal2008a,vidal2010,chen2010} and hence 
accurately reproduces correlators across a wide range of length scales\citestop{vidal2007,vidal2008a,giovannetti2008,pfeifer2009,evenbly2009} The development of a general formalism for anyonic tensor networks is therefore also advantageous for the study of {1-D} anyonic systems.

This paper describes how any tensor network algorithm may be adapted to systems of anyons in one or two dimensions using structures which explicitly implement the quantum group symmetry of the anyon model. As a specific example we demonstrate the construction of the anyonic {1-D} MERA, which we then apply to an infinite chain of interacting Fibonacci anyons at criticality.
The approach which we present is 
completely general, and can be applied to any species of anyons and any tensor network ansatz.

\section{Anyonic states\label{sec:anyonstates}}

\begin{figure}
\includegraphics[width=246.0pt]{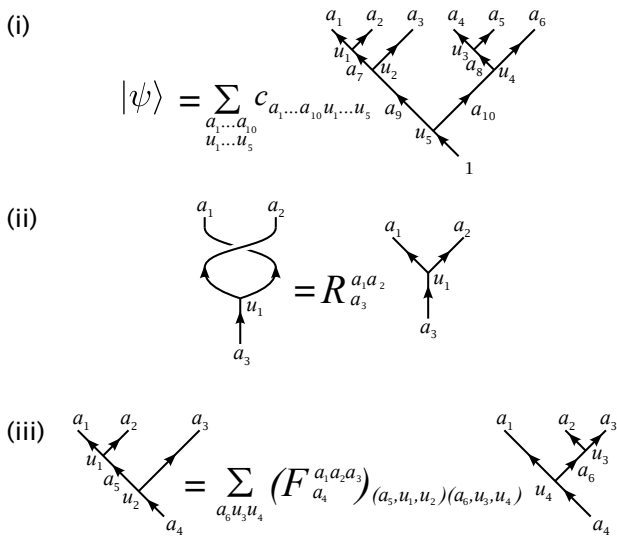}
\caption{(i)~Example representation of a state $|\psi\ra$ in a fusion tree basis for a system of $6$ anyons. Labels $a_i$ indicate charges associated with edges of the fusion tree graph, and labels $u_i$ are degeneracies associated with vertices. The structure of the tree corresponds to a choice of basis, and does not affect the physical content of the theory.
(ii)~Braiding may be used to change the ordering of the leaves of a fusion tree basis, or to represent anyon exchange.
(iii)~$F$-moves convert between the bases associated with different fusion trees.
\label{fig:anyons}}
\end{figure}

Consider a lattice $\mc{L}_0$ of $n$ sites populated by anyons. In contrast to bosonic and fermionic systems, for many anyon models the total Hilbert space $\mbb{V}_{\mc{L}_0}$ can not be divided into a tensor product of local Hilbert spaces. Instead, a basis is defined by introducing a specific fusion tree (e.g. \fref{fig:anyons}(i)). The fusion tree is always constructed on a linear ordering of 
anyons, and while the {1-D} lattice naturally exhibits such an ordering, for 2-D lattices some linear ordering 
must be imposed.
Each line is then labelled with a charge index $a_i$ such that the labels are consistent with the fusion rules of the anyon model,
\begin{equation}
a\times b\rightarrow \sum_{c} N_{ab}^{c} \,c.\label{eq:fusion}
\end{equation}
For anyon types where some entries of the multiplicity tensor $N_{ab}^c$ take values greater than 1, a label $u_i$ is also affixed to the vertex which represents the fusion process to distinguish between the different copies of charge $c$.
The edges of the graph which are connected to a vertex only at their lower end are termed ``leaves'' of the fusion tree, and we will associate these leaves with the charge labels $a_1\ldots a_n$. Different orderings of the leaves on a fusion tree may be interconverted by means of braiding (\fref{fig:anyons}(ii)), and different fusion trees, corresponding to different bases of states, may be interconverted by means of $F$ moves (\fref{fig:anyons}(iii))\citestop{kitaev2006,bonderson2007}
In some situations it may also be useful to associate a further index $b_i$ 
with each of the leaves of the fusion tree. For example, if the leaves are equated with the sites of a physical lattice, then this additional index may be used to enumerate additional non-anyonic degrees of freedom associated with that lattice. 
For simplicity we will usually leave 
these extra indices $b_1\ldots b_n$ implicit, as we have done in \fref{fig:anyons}, 
as they do not directly participate in anyonic manipulations such as $F$ moves and braiding.

Let the total number of charge labels on the fusion tree be given by $m$, where $m\geq n$. 
For abelian anyons the fusion rules uniquely constrain all $a_i$ for $i>n$, and provided there are no constraints on the total charge, the total Hilbert space reduces to a product of local Hilbert spaces $\mbb{V}$, such that $\mbb{V}_{\mc{L}_0} = \mbb{V}^{\otimes n}$. For nonabelian anyons, additional degrees of freedom arise because some fusion rules admit multiple outcomes, permitting certain $a_i\ (i>n)$ to take on multiple values while remaining consistent with the fusion rules, and the resulting Hilbert space does not necessarily admit a tensor product structure.

We will now associate a parameter $\nu_{i,a_i}$ with each charge on the fusion tree, which we will term the degeneracy. This parameter corresponds to the number of possible fusion processes by which charge $a_i$ may be obtained at location $i$. Where charge $a_k$ arises from the fusion of charges $a_i$ and $a_j$, then $\nu_{k,a_k}$ will satisfy
\begin{equation}
\nu_{k,a_k} = \sum_{a_i,a_j} \nu_{i,a_i}\nu_{j,a_j} N^{a_k}_{a_i a_j}.\label{eq:compounddegens}
\end{equation}
For systems where the only degrees of freedom are anyonic, degeneracies on the physical lattice $\mc{L}_0$ (i.e. $\nu_{i,a_i}$, $1\leq i\leq n$) will take values of 0 or 1 depending on whether a charge $a_i$ is permitted on lattice site $i$. Higher values of $\nu_{i,a_i}$ may be used on the physical lattice 
if there is also a need to represent additional 
non-anyonic degrees of freedom, 
enumerated by indices $b_1\ldots b_n$.

\begin{figure}
\includegraphics[width=246.0pt]{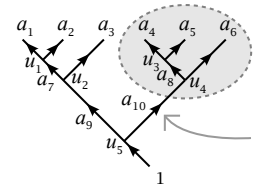}
\caption{
The leaves of this fusion tree carry the charge labels
$a_1$ to $a_6$. 
An edge which is not a leaf, labelled with charge $a_{10}$, is indicated by the large grey arrow. 
The portion of the fusion tree extending from edge $a_{10}$ out to the leaves
is indicated by the grey ellipse. If a degeneracy index $\mu_{10}$ is associated with charge $a_{10}$, then for a given value of $a_{10}$, index $\mu_{10}$ will 
enumerate all compatible labellings of the highlighted portion of the fusion tree.\label{fig:exampleedges}}
\end{figure}

\begin{figure}
\includegraphics[width=246.0pt]{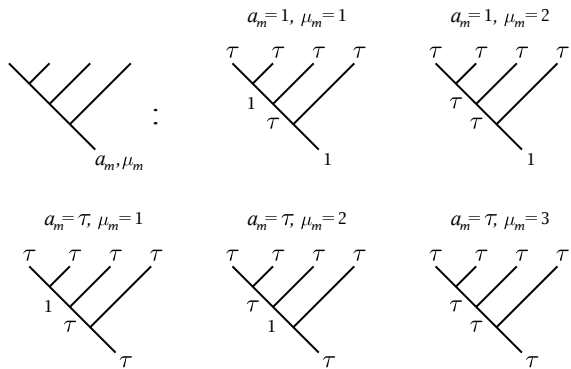}
\caption{Example enumeration of states according to $a_m$ and $\mu_m$ for a fusion tree describing 
four Fibonacci anyons. The Fibonacci anyon model has one non-vacuum charge label ($\tau$) and one non-trivial fusion rule, $\tau\times\tau\rightarrow 1+\tau$. Because the charges $1$ and $\tau$ are both self-dual, no arrows are required on diagrammatic representations of Fibonacci anyon fusion trees.\label{fig:exampleenum}}
\end{figure}

Up to this point we have parameterised our Hilbert space in terms of explicit labellings of the fusion tree. We now adopt a different approach: Consider an edge $i$ of the fusion tree which is not a ``leaf''. 
As well as labelling this edge with a charge $a_i$ we may introduce a second index $\mu_i$, running from 1 to $\nu_{i,a_i}$. Each pair of values $\{a_i,\mu_i\}$ may be associated with a unique charge labelling for the portion of the fusion tree 
from edge $i$ out to the leaves,
with these labellings being compatible with the fusion rules in the presence of a charge of $a_i$ on site $i$ (for an illustration of this, see \fref{fig:exampleedges}). 
Provided we know the structure of the fusion tree above $i$ and have a systematic means of associating labellings of that portion of the tree with values of $\mu_i$, then in lieu of stating the values of all $a_j$ for edges $j$ involved in that portion of the tree, we may simply specify the value of the degeneracy index $\mu_i$. 
In this way we may specify an entire state in the form
\begin{equation}
|\psi\ra = \sum_{\mu_{m}} c_{a_{m}\mu_{m}} |a_{m},\mu_{m}\ra\label{eq:statepsi_pre}
\end{equation}
where $a_{m}$ is the total charge obtained on fusing all the anyons. 
The index $\mu_{m}$, which is the degeneracy index associated with the total charge of the fusion tree, may be understood as systematically enumerating all possible labellings of the entire fusion tree including charge labels, vertex labels, and any labels associated with additional non-anyonic degrees of freedom. 
For an example, see \fref{fig:exampleenum}.
Note that for a given edge $i$, the value of the degeneracy $\nu_{i,a_i}$ may vary with the charge $a_i$ and consequently the range of the degeneracy index $\mu_{m}$ in Eq.~\eref{eq:statepsi_pre} is dependent on the value of the charge $a_{m}$.

The notation of Eq.~\eref{eq:statepsi_pre} should be contrasted with that of
\fref{fig:anyons}(i). In the latter, the number of indices on $c$ depends upon the number of charge labels on the fusion tree, whereas in the former, the tensor describing the state is always indexed by just one pair of labels---charge and degeneracy---which will prove advantageous in constructing a tensor network formalism for systems of anyons.

We now choose to restrict our attention to systems having the identity charge. We may do this without loss of generality as a state on $n$ lattice sites with a total charge $a_{m}$ may always be equivalently represented by a state on $n+1$ lattice sites whose total charge is the identity, with a charge $\overline{a_{m}}$ on lattice site $n+1$. This additional charge annihilates the total charge $a_{m}$ of sites $1\ldots n$ to give the vacuum. The expression for $|\psi\ra$ then becomes
\begin{equation}
|\psi\ra = \sum_{\mu_{m'}} c_{1\mu_{m'}} |1,\mu_{m'}\ra\label{eq:statepsi}
\end{equation}
where $\mu_{m'}$ ranges from 1 to the dimension of the Hilbert space of the system of $n$ sites with total charge $a_m$.
Consequently we may represent the state $|\psi\ra$ of a system of anyons by means of the vector $c_{1\mu_{m'}}$. For simplicity of notation, we will take greek indices from the beginning of the alphabet to correspond to pairs of indices $\{a_i,\mu_i\}$ consisting of a charge index and the associated degeneracy index. The vector $c_{1\mu_{m'}}$ will therefore be denoted simply $c^\alpha$, with the understanding that in this case 
the charge component $a_{m'}$ of multi-index $\alpha$ takes 
only
the value 1. (Multi-index $\alpha$ is raised as we will shortly introduce a diagrammatic formalism in which vector $c$ is represented by an object with a single upward-going leg. In this formalism, upward- and downward-going legs may be associated with upper and lower multi-indices respectively.)


\section{Anyonic operators}

We will divide our consideration of anyonic operators into two parts. First we shall consider operators which map a state on some Hilbert space $\mc{H}$ into another state on the same Hilbert space. When applied to a state represented by $c^\alpha$, such an operator leaves the degeneracies of the charges in multi-index $\alpha$ unchanged. We will therefore call these
\emph{degeneracy-preserving} anyonic operators.
Then we will consider those operators which map a state on some Hilbert space $\mc{H}$ into a state on some other Hilbert space $\mc{H}'$. These operators may represent processes which modify the environment, for example by adding or removing lattice sites, and also play an important part in anyonic tensor networks, for instance taking the role of isometries in the TTN and MERA. 
As these operators can change the degeneracies of charges in a multi-index $\alpha$, we will call them \emph{degeneracy-changing} anyonic operators.
More generally, the degeneracy-preserving anyonic operators may be considered a subclass of the degeneracy-changing anyonic operators for which $\mc{H}=\mc{H}'$.

\subsection{Degeneracy preserving anyonic operators\label{sec:anyonops}}

We begin with those operators which map
states on some Hilbert space $\mc{H}$ into other states on the same Hilbert space $\mc{H}$. 
Examples of these operators include Hamiltonians, reduced density matrices, and unitary transformations such as the disentanglers of the MERA.

\begin{figure}
\includegraphics[width=246.0pt]{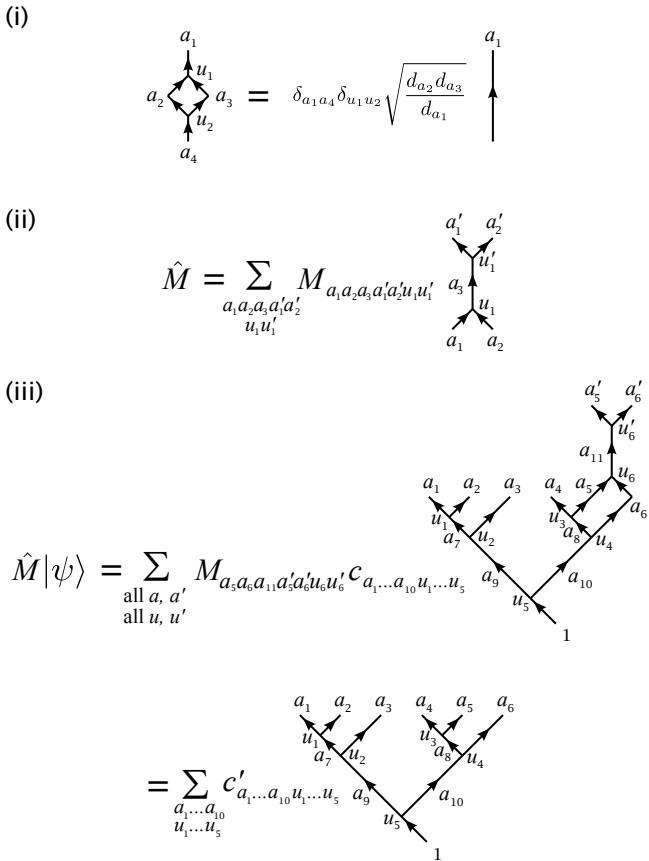}
\caption{
(i) Loops are eliminated by replacing them 
with an equivalent numerical factor determined by the normalisation convention. 
The factor given here corresponds
to the diagrammatic isotopy convention employed in \protect{\rcite{bonderson2007}}.
(ii) Definition of a simple two-site anyonic operator. 
(iii) Application of an operator to a state is performed by connecting the diagrams' free legs. By performing $F$ moves and eliminating loops (and in more complex examples, also braiding) it is possible to obtain an expression for the resulting state in the original basis. 
\label{fig:operators}}
\end{figure}

First, we introduce splitting trees. The space of splitting trees is dual to the space of fusion trees. While the space of fusion trees consists of labelled directed graphs whose number of branches increases monotonically when read from bottom to top, the space of splitting trees consists of labelled directed graphs whose number of branches increases monotonically when read from top to bottom. 
An inner product is defined by connecting the leaves of fusion and splitting trees which have equivalent linear orderings of the leaves (
braiding first 
if necessary), 
then eliminating all loops as per \fref{fig:operators}(i), with $F$ moves performed as required. 
%

Anyonic operators
may always be written as a sum over fusion and splitting trees, such as the two-site operator $\hat M$
shown in \fref{fig:operators}(ii), and for degeneracy-preserving anyonic operators it is 
always possible to choose the splitting tree to be the adjoint of the fusion tree.
To apply an operator 
to a state the two corresponding diagrammatic representations are connected as shown in \fref{fig:operators}(iii), and closed loops may be eliminated as shown in \fref{fig:operators}(i). 
Sequences of $F$ moves, braiding, and loop eliminations may be performed until the diagram has been reduced once more to a fusion tree without loops on a lattice of $n$ sites.

Much as the state of an anyonic system may be represented by a vector $c^\alpha$, anyonic operators may be represented by a matrix $\Mop$. Each value of $\alpha$ corresponds to a pair $\{a_i,\mu_i\}$ where $a_i$ is a possible charge of the central edge of the operator diagram (e.g. $a_3$ in \fref{fig:operators}(ii)), and $\mu_i$ is a value of the degeneracy index associated with charge $a_i$. We will denote the degeneracy of $a_i$ by $\nu_{a_i}$. Similarly, values of $\beta$ correspond to pairs $\{a_j,\mu_j\}$ where $a_j$ has degeneracy $\nu_{a_j}$.
For degeneracy-preserving anyonic operators the charge indices
$a_i$ and $a_j$ necessarily take on the same range of values, and $\nu_{a_i}=\nu_{a_j}$ when $a_i=a_j$. The values of $\nu_{a_i}$ may equivalently be calculated from either the fusion tree making up the top half or the splitting tree making up the bottom half of the operator diagram.

\begin{figure}
\includegraphics[width=246.0pt]{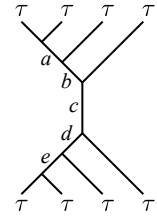}
\caption{An operator acting on four Fibonacci anyons. The values of the coefficients $M_{abcde}$ may be specified as a block-diagonal matrix $M_\alpha^{~\beta}$, for example as in \protect{\tref{tab:examplematrix}}.\label{fig:exampleoperator}}
\end{figure}

A well-defined anyonic operator $\hat M$ must respect the (quantum) symmetry group of the anyon model, and consequently all entries in $\Mop$ for which $a_i\not=a_j$ will be zero. However, in contrast with $c^\alpha$ we do not require that $a_i=a_j=1$. When $\hat M$ is a degeneracy-preserving operator, matrix $\Mop$ is therefore a square matrix of side length
\begin{equation}
\ell_M=\sum_{a_i} \nu_{a_i},
\end{equation}
which may be organised to exhibit a structure which is block diagonal in the charge indices $a_i$ and $a_j$, and for which the blocks are also square. 
As an example consider \fref{fig:exampleoperator}, which shows an operator acting on four Fibonacci anyons. An example matrix $\Mop$ for an operator of this form is given in \tref{tab:examplematrix}, from which the entries of $M_{abcde}$ can be reconstructed, e.g. $M_{\tau 1 \tau 1 \tau}=3$.

\begin{table}
\begin{equation*}
\Mop=\quad\begin{array}{cc|ccccccc}
&&&\multicolumn{5}{c}{a_j,\mu_j}&\\
&&&1,1&1,2&\tau,1&\tau,2&\tau,3&\\
\hline
\multirow{5}{*}{$a_i$,$\mu_i$}&1,1&
\multirow{5}{*}{$\left(\begin{array}{c}\!\\ \!\\ \!\\ \!\\ \!\end{array}\right.$}& 1&0.5&0&0&0 &
\multirow{5}{*}{$\left.\begin{array}{c}\!\\ \!\\ \!\\ \!\\ \!\end{array}\right)$}\\
&1,2&& 0.5&1&0&0&0 &\\
&\tau,1&& 0&0&1&2&-1 &\\
&\tau,2&& 0&0&2&3&-1 &\\
&\tau,3&& 0&0&1&1&1 &
\end{array}
\end{equation*}
~

\begin{tabular}{|cc|ccccc|}
\hline
$a_i$&$\mu_i$&~&$a$&$b$&$c$&\\
\hline\hline
1&1&&1&$\tau$&1&\\
1&2&&$\tau$&$\tau$&1&\\
$\tau$&1&&1&$\tau$&$\tau$&\\
$\tau$&2&&$\tau$&1&$\tau$&\\
$\tau$&3&&$\tau$&$\tau$&$\tau$&\\
\hline
\end{tabular}
~~~~
\begin{tabular}{|cc|ccccc|}
\hline
$a_j$&$\mu_j$&~&$e$&$d$&$c$&\\
\hline\hline
1&1&&1&$\tau$&1&\\
1&2&&$\tau$&$\tau$&1&\\
$\tau$&1&&1&$\tau$&$\tau$&\\
$\tau$&2&&$\tau$&1&$\tau$&\\
$\tau$&3&&$\tau$&$\tau$&$\tau$&\\
\hline
\end{tabular}
\caption{Matrix representation $M_\alpha^{~\beta}$ for an example operator of the form shown in \protect{\fref{fig:exampleoperator}}. Multi-index $\alpha$ corresponds to index pair $\{a_i,\mu_i\}$ and multi-index $\beta$ corresponds to pair $\{a_j,\mu_j\}$. Subject to an appropriate ordering convention for $\mu_i$ and $\mu_j$, these indices may be related to the fusion tree labels $a,b,c,d,e$ of \protect{\fref{fig:exampleoperator}} as shown. Note that as $c$ is the charge on the central leg of \protect{\fref{fig:exampleoperator}}, all nonzero entries of $M_\alpha^{~\beta}$ satisfy $a_i=a_j=c$.\label{tab:examplematrix}}
\end{table}


\subsection{Degeneracy changing anyonic operators\label{sec:degenexp}}

We now introduce the second class of anyonic operators, which map states in some Hilbert space $\mc{H}$ into some other Hilbert space $\mc{H}'$. These operators may reduce or increase the degeneracy of any charge present in the spaces on which they act, and may even project out entire charge sectors by setting their degeneracy to zero.
When these operators are written in the conventional notation of \fref{fig:operators}, the fusion and splitting trees will not be identical. Further, we may choose to allow combinations of degeneracies which do not naturally admit complete decomposition into individual anyons. For example, a degeneracy-changing operator may map a state on five Fibonacci anyons (having total degeneracies $\nu_1=3$, $\nu_\tau=5$) into a state having degeneracies $\nu_1=2$, $\nu_\tau=2$. As these degeneracies do not admit decomposition into an integer number of nondegenerate anyons, it is necessary 
to associate an index $u_i$ with the single open leg of the fusion tree. This index behaves identically to the vertex indices $u_i$ of \fref{fig:anyons}, serving to enumerate the different copies of each individual charge, and as with the vertex indices of \fref{fig:anyons}, it is absorbed into the degeneracy index $\mu_i$.

As a further example, a state having degeneracies $\nu_1=4$, $\nu_\tau=4$ could be associated with a fusion tree having either one leg, or two legs each with degeneracies $\nu_1=0$, $\nu_\tau=2$. Again, indices $u_i$ would have to be associated with each open leg.




Matrix representations of degeneracy-changing anyonic operators may also 
be constructed, and when they 
are written in block diagonal form, the matrices and their blocks may be rectangular rather than square. Degeneracy-changing anyonic operators therefore 
represent a generalisation of the degeneracy-preserving anyonic operators discussed in \sref{sec:anyonops}. It is worth noting that the presence of indices $u_i$ on the open legs of the fusion or splitting trees of an operator do not automatically imply that it is a degeneracy-changing anyonic operator: The defining characteristic of a degeneracy-preserving anyonic operator is that it maps a state in a Hilbert space $\mc{H}$ into a state in the same Hilbert space $\mc{H}$, and consequently both the matrix as a whole and all of its blocks are square. Thus a degeneracy-preserving anyonic operator may act on states having additional indices $u_i$ on their open legs, and the resulting state may be expressed in the form of the same fusion tree, with the same additional indices on the open legs.



Operators which change degeneracies may 
represent physical processes which change the accessible Hilbert space of a system. 
As we will see in \sref{sec:MERAconstr}, 
they may also be used in tensor network algorithms as part of an efficient representation of particular states or subspaces of a Hilbert space, for example the ground state or the low energy sector of a local Hamiltonian.

\begin{figure}
\includegraphics[width=246.0pt]{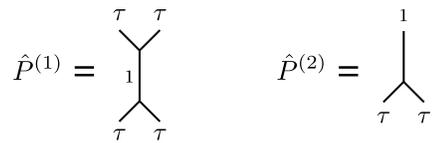}
\caption{Diagrammatic representation of operators $\hat P^{(1)}$ \eref{eq:P1} and $\hat P^{(2)}$ \eref{eq:P2}. The charges on all leaves are non-degenerate.\label{fig:projectionops}}
\end{figure}

This distinction between degeneracy-changing and degeneracy-preserving anyonic operators is 
clearly seen with a simple example. Let $|\psi\ra$ be a state on six Fibonacci anyons. This state can be parameterised by a vector $c^\alpha$, which has five components. We now define two projection operators, $\hat P^{(1)}$ and $\hat P^{(2)}$ (\fref{fig:projectionops}), each of which acts on the fusion space of anyons $\tau_1$ and $\tau_2$. 
Operator $\hat P^{(1)}$ is degeneracy-preserving, and projects $c^\alpha$ into the subspace in which anyons $\tau_1$ and $\tau_2$ fuse to the identity. Its matrix representation is
\begin{equation}
P_{\p{(1)}\,\alpha}^{(1)\,\p{\alpha}\beta} =
\left(
\begin{array}{cc}
1&0\\0&0
\end{array}
\right)\label{eq:P1}
\end{equation}
where the first value of each multi-index corresponds to a charge of 1, and the second to a charge of $\tau$.
Operator $\hat P^{(2)}$ performs the same projection, but is degeneracy-changing. Its matrix representation is written
\begin{equation}
P_{\p{(2)}\,\alpha}^{(2)\p{\,\alpha}\beta}=(~1~0~).
\label{eq:P2}
\end{equation}
Both operators perform equivalent projections, in the sense that 
\begin{equation}
\la\psi|\hat P^{(1)\dagger}\hat P^{(1)}|\psi\ra = \la\psi|\hat P^{(2)\dagger}\hat P^{(2)}|\psi\ra.
\end{equation} 
When $\hat P^{(1)}$ acts on $|\psi\ra$ it leaves the Hilbert space unchanged, and hence the vector $c'^\alpha$ describing state $|\psi'\ra=\hat P^{(1)}|\psi\ra$ is once again a five-component vector, although in an appropriate basis some components will now necessarily be zero. In contrast $\hat P^{(2)}$ explicitly reduces the dimension of the Hilbert space, and the vector $c''^\alpha$ describing state $|\psi''\ra=\hat P^{(2)}|\psi\ra$ is of length two, describing a fusion tree on only four Fibonacci anyons (as both $\tau_1$ and $\tau_2$ have been eliminated). One consequence of this distinction is that while $(\hat P^{(1)})^2=\hat P^{(1)}$, the value of $(\hat P^{(2)})^2$ is undefined.

\section{Anyonic tensor networks}

\subsection{Diagrammatic notation\label{sec:tensordiagrammatic}}

The diagrammatic notation conventionally employed in the study of anyonic systems, and used here in Figs.~\ref{fig:anyons} and \ref{fig:operators}, is well suited to the complete description of anyonic systems, as it provides a physically meaningful depiction of the entire Hilbert space. 
However, the number of parameters required for such a description grows exponentially in the system size, and because it is necessary to explicitly assign every index to a specific charge or degeneracy, specification of a tensor network rapidly becomes inconveniently verbose (for example see \fref{fig:operators}(iii)). 

\begin{figure}
\includegraphics[width=246.0pt]{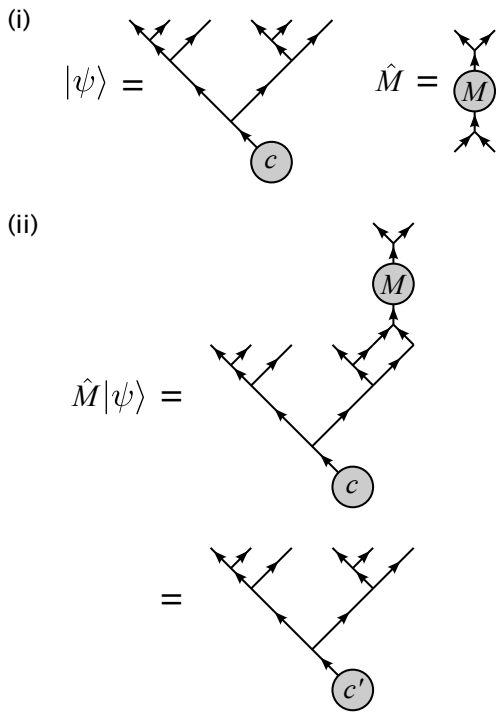}
\caption{(i) Diagrammatic representation of a state $|\psi\ra$ and two-site operator $\hat M$ expressed in terms of degeneracy indices. (ii) Application of $\hat M$ to state $|\psi\ra$. Grey shapes represent tensors with charge and degeneracy multi-indices, with each leg of the shape corresponding to one charge and degeneracy index pair. These diagrams represent the same state, operator, and process as \protect{\fref{fig:anyons}}(i) and \protect{\fref{fig:operators}}(ii)-(iii).\label{fig:blobtensors}}
\end{figure}

In the preceding Sections, we developed techniques whereby anyonic states and operators could be 
represented as vectors and matrices, bearing only one or two multi-indices apiece. We now introduce the graphical notation which complements this description, and in which we will formulate anyonic tensor networks. \fref{fig:blobtensors}(i) gives the graphical representations of a state $|\psi\ra$ associated with a vector $c^\alpha$, and of an operator $\hat M$ associated with a matrix $\Mop$.
The circle marked $c$ corresponds to the vector $c^\alpha$, and the circle marked $M$ corresponds to the matrix $\Mop$. In general, grey circles correspond to tensors, and the number of legs on the circle corresponds to the number of multi-indices on the associated tensor. Each multi-index is also associated with a fusion or splitting tree structure, which is specified graphically. For reasons to be discussed shortly, we will require that no tensor ever have more than three multi-indices.
As the legs of the grey shapes are each associated with a multi-index, they carry both degeneracy and charge indices. Consequently it is not necessary to explicitly assign labels to the fusion/splitting trees, as these labellings are contained implicitly in the degeneracy index (for example see \tref{tab:examplematrix}, where specifying the values of $\{a_i,\mu_i\}$ and $\{a_j,\mu_j\}$ is equivalent to fully labelling the fusion and splitting trees of \fref{fig:exampleoperator}). 

The fusion or splitting tree associated with a particular multi-index may be manipulated in the usual way by means of braids and $F$ moves, recalling that each component of the tensor is associated with a particular labelling of the fusion and splitting trees via the corresponding values of the multi-indices. 
Manipulations performed upon a particular tree thus generate unitary matrices which act upon the multi-index that corresponds to the labellings of that particular tree.

The application of an operator to a state is, unsurprisingly, performed by connecting the appropriate diagrams, as shown in \fref{fig:blobtensors}(ii). For operators of the type discussed in \sref{sec:anyonops}, the outcome is necessarily a new state in the same Hilbert space, which consequently can be described by a new state vector $c'^\alpha$, as shown. However, in general an operator $\hat M$ will not act on the entire Hilbert space of the system, and so will be described by a tensor constructed on the fusion space of some subset of lattice sites, and not on the system as a whole. Operator $\hat M$ acting on state $|\psi\ra$ in \fref{fig:blobtensors}(ii) is an example of this. Because $c^\alpha$ describes a six-site system but $\Mop$ is constructed on the fusion space of two sites, 
the multi-indices of $\Mop$ span a significantly smaller Hilbert space than that of $c^\alpha$ and we cannot simply write 
\begin{equation}
c'^\beta = c^{\alpha}M_\alpha^{\phantom{\alpha}\beta}
\end{equation}
(using Einstein notation, where repeated multi-indices are assumed to be summed).
Instead, we must understand how to expand the matrix representation of an operator on some number of sites $x$, to obtain its matrix representation as an operator on $x'$ sites, where $x'>x$. 



\subsection{Site expansion of anyonic operators\label{sec:siteexpanyop}}

The multiplicity tensor $N^{c}_{ab}$ describes the fusion of two charges without degeneracies. It is easily extended to incorporate degeneracies of the charges, 
and we will denote this expanded multiplicity tensor $\tilde N^\gamma_{\alpha\beta u}$ 
where 
multi-indices $\alpha$, $\beta$, and $\gamma$ are associated with the pairs 
$\{a,\mu_a\}$, $\{b,\mu_b\}$, and $\{c,\mu_c\}$ respectively, and for given values of $\alpha$, $\beta$, and $\gamma$, $u$ runs from 1 to $N^c_{ab}$. 
The degeneracies associated with charges $a$, $b$, and $c$ are denoted $\nu_{a}$, $\nu_{b}$, and $\nu_c$ respectively. As with $\mu_a$, $\mu_b$, and $\mu_c$, there is an implicit additional index on each degeneracy $\nu_x$ representing the edge of the tree on which 
charge $x$ resides. 
The values of $\nu_a$ and $\nu_b$ may be chosen arbitrarily (for example, $\nu_{a}|_{a=1}$ may differ from $\nu_{b}|_{b=1}$), but the degeneracies associated with the values of $c$ must satisfy
\begin{equation}
\nu_{c}=\sum_{a,b} \nu_{a}\nu_{b}N^{c}_{ab}
\end{equation}
in accordance with Eq.~\eref{eq:compounddegens}. When this constraint is satisfied, 
every quadruplet of indices $\{a,\mu_a,b,\mu_b\}$ corresponding to a unique pair of choices for $\alpha$ and $\beta$ may be associated with $N^c_{ab}$ distinct
pairs of indices $\{c,\mu_c\}$ for each $c\in a\times b$. These pairs $\{c,\mu_c\}$ are enumerated by the additional index $u$. 
This defines a 1:1 mapping between sets of values on $\{a,\mu_a,b,\mu_b,u\}$ and pairs $\{c,\mu_c\}$,
%
and we set the corresponding entries in $\tilde N^\gamma_{\alpha\beta u}$ to 1, with all other entries being zero.
A simple example is given in \tref{tab:exampleN}.
\begin{table}
\begin{tabular}{c|c}
Pair & Assigned pentuplet \\
\{$c,\mu_c$\} & \{$a,\mu_a,b,\mu_b,u$\} \\
\hline
$1,~1$ & $1,~1,~1,~1,~1$ \\
$1,~2$ & $\tau,~1,~\tau,~1,~1$ \\
$1,~3$ & $\tau,~1,~\tau,~2,~1$ \\
$\tau,~1$ & $1,~1,~\tau,~1,~1$ \\
$\tau,~2$ & $1,~1,~\tau,~2,~1$ \\
$\tau,~3$ & $\tau,~1,~1,~1,~1$ \\
$\tau,~4$ & $\tau,~1,~\tau,~1,~1$ \\
$\tau,~5$ & $\tau,~1,~\tau,~2,~1$
\end{tabular}
\caption{Construction of $\tilde N^\gamma_{\alpha\beta u}$ for a fusion vertex for Fibonacci anyons.
In this example $a$ may take charges $1$ and $\tau$ each with degeneracy 1, and $b$ may take charges $1$ and $\tau$ with degeneracies 1 and 2 respectively. By Eq.~\protect{\eref{eq:compounddegens}}, charge $c$ may therefore take values $1$ and $\tau$ with degeneracies 3 and 5 respectively.
A correspondence between the values of multi-index $\gamma$ and of multi-indices $\alpha$ and $\beta$ is established in some systematic manner, with each assignation satisfying $c\in a\times b$, and for Fibonacci anyons the index $u$ is trivial as all multiplicities $N^c_{ab}$ are zero or one.
An example 
assignation is shown in the table. The corresponding entries of $\tilde N$ are then set to 
1, with all other entries zero. For example, the fourth row indicates that $\tilde N^{(\tau,1)}_{(1,1)(\tau,1)1}=1$.\label{tab:exampleN}}
\end{table}



By virtue of their derivation from $N^{c}_{ab}$, the object $\tilde N^\gamma_{\alpha\beta u}$ and its conjugate $\tilde N^{\dagger\alpha\beta u}_{\phantom{\dagger}\gamma}$ 
represent application of the anyonic fusion rules, and may be associated with vertices of the splitting and fusion trees. Under the isotopy invariance convention there is an additional factor of $[d_{c}/(d_{a}d_{b})]^\frac{1}{4}$ associated with the fusion of 
charges $a$ and $b$ into $c$, where $d_{x}$ is the quantum dimension of charge $x$, and similarly for splitting, but we will account for these factors separately. Thus constructed, the tensors $\tilde N$ satisfy $\tilde N^\gamma_{\alpha\beta u}\tilde N^{\dagger\alpha\beta u}_{\p\dagger\epsilon}=\delta_\epsilon^{\p\epsilon\gamma}$.



When used as a representation of the fusion rules, the generalised multiplicity tensor $\tilde N^\gamma_{\alpha\beta u}$ and its conjugate $\tilde N^{\dagger\alpha\beta u}_{\phantom{\dagger}\gamma}$ permit us to increase or decrease the number of multi-indices on a tensor in a manner which is consistent with the fusion rules of the quantum symmetry group. This process is reversible provided the symmetry group is abelian or, for a nonabelian symmetry group, provided the total number of multi-indices on the tensor does not at any time exceed three.
In constructing and manipulating a tensor network for a system of anyons, we will require only objects which respect the fusion rules of the anyon model. It is a defining property of such objects that when the number of multi-indices they possess is reduced to 1 by repeated application of $\tilde N$ and $\tilde N^\dagger$, non-zero entries may be found only in the vacuum sector. We imposed this requirement for states in \sref{sec:anyonstates}, and it is equivalent to the restriction we imposed on anyonic operators in \sref{sec:anyonops}.
In \rcite{singh2009} an equivalent condition was observed for tensors remaining unchanged under the action of a Lie group, and these tensors were termed \emph{invariant}. When working with invariant tensors, we may separately evaluate the components of the tensors acting on the degeneracy spaces (e.g. the nonzero blocks of $M_\alpha^{\phantom{\alpha}\beta}$), and the factors arising from loops and vertices of the associated spin network. This property greatly simplifies the contraction of pairs of tensors. 

\begin{figure}
\includegraphics[width=246.0pt]{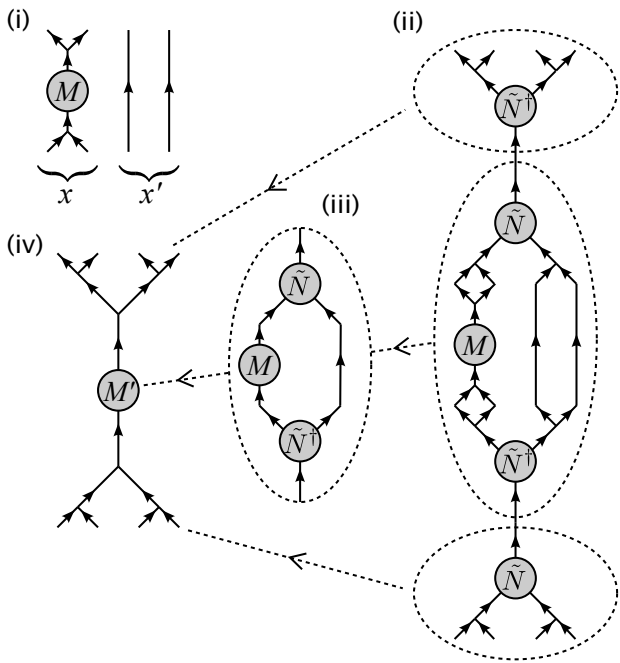}
\caption{``Raising'' of an operator $\hat M$ from sites $x$ to sites $x+x'$: (i) Operator $\hat M$ defined only on sites denoted $x$. (ii) Resolutions of the identity are inserted above and below $\hat M$, being constructed from tensors $\tilde N$ and $\tilde N^\dagger$. The central portion of this diagram is identified as corresponding to the new matrix $M'^{~\beta}_\alpha$ which describes $\hat M$ on $x+x'$. (iii) Loop and vertex factors in the central region are evaluated separately and eliminated. (iv) The tensor network corresponding to the new central portion is contracted. The $\tilde N$ and $\tilde N^\dagger$ tensors outside the central region become vertices of the fusion and splitting trees associated with $M'^{~\beta}_\alpha$. Together the trees and the matrix $M'^{~\beta}_\alpha$ constitute the raised version of $\hat M$. \label{fig:raiseoperator}}
\end{figure}

In addition to increasing or decreasing the number of legs of a tensor, we may also use $\tilde N$ to ``raise'' the matrix representation of an operator from the space of $x$ sites to the space of $(x+x')$ sites. This is shown in \fref{fig:raiseoperator}, and the matrix representation of the raised operator is given by
\begin{equation}
M_{\p{\prime}\alpha}^{\prime\p{\alpha}\beta} = M_\gamma^{\p{\alpha}\delta}\tilde N^{\dagger \gamma\epsilon u}_{\p{\dagger}\alpha} \tilde N^\beta_{\delta\epsilon u}\label{eq:Mprime}
\end{equation}
where multi-index $\epsilon$ describes the fusion space of all sites in $(x+x')$ but not in $x$. 
Because the numeric factors associated with loops, vertices, and braiding (where applicable) are handled separately,
no factors of quantum dimensions appear in Eq.~\eref{eq:Mprime}.

To act an operator $\hat M$ on a state $|\psi\ra$ in the matrix representation, we therefore connect the diagrams for $\hat M$ and $|\psi\ra$, eliminate all loops, and then raise the  matrix representation of the operator $\hat M$ using Eq.~\eref{eq:Mprime}, repeatedly if necessary, until the resulting matrix $M_{\p{\prime}\alpha}^{\prime\p{\alpha}\beta}$ may be applied directly to the state vector $c^\alpha$. 
Similarly it is possible to combine the matrix representations of operators, by connecting their diagrams appropriately, eliminating loops, and performing any required raising so that both operators act on the same fusion space. Their matrix representations can then be combined to yield the matrix representation of the new operator:
\begin{equation}
M^{(1\times2)\p{\alpha}\beta}_{\p{(1\times2)}\alpha} = M^{(1)\p{\alpha}\gamma}_{\p{(1)}\alpha} M^{(2)\p{\gamma}\beta}_{\p{(2)}\gamma},\label{eq:contractmatrices}
\end{equation}
and the fusion/splitting tree associated with this new operator is obtained as shown in \fref{fig:raiseoperator}.

Note that as yet, we have not described how two objects may be combined if their multi-indices are both up or both down, and are connected by a curved line. To contract such objects together, it is necessary to understand how bends act on the central matrix of an operator. Once this is understood, the bend can be absorbed into one of the central matrices, so that the connection is once again between an upper and a lower multi-index as in Eq.~\eref{eq:contractmatrices}. This process is described in \sref{sec:anyonmanip}.

\subsection{Manipulation of anyonic operators\label{sec:anyonmanip}}

\begin{figure}
\includegraphics[width=246.0pt]{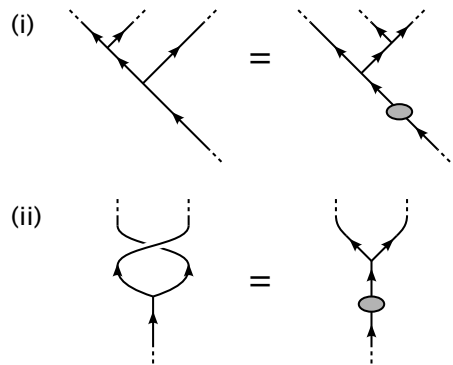}
\caption{(i) $F$ move, and (ii) braiding, performed on a section of fusion tree in the diagrammatic notation of \protect{\sref{sec:tensordiagrammatic}}.\label{fig:anyonopmanip1}}
\end{figure}

As observed in \sref{sec:siteexpanyop}, when we describe a system entirely in terms of objects invariant under the action of the symmetry group, we may account separately for the numerical normalisation factors associated with the spin network. 
However, as well as affecting these numerical factors, transformations of the fusion or splitting tree of an anyonic operator will typically also generate unitary matrices which act on the matrix representation of the operator. These matrices respect the symmetry of the anyon model, and thus can be written as block-diagonal matrices where each block is 
a unitary matrix acting on 
a particular charge sector. 
In terms of the diagrammatic notation of \sref{sec:tensordiagrammatic}, $F$ moves and braids therefore result in the insertion of a unitary matrix, as shown in \fref{fig:anyonopmanip1}. These matrices, whose entries are derived from the tensors $(F^{abc}_{d})_{(euv)(fu'v')}$ and $R^{ab}_c$ respectively, are raised if required, as described in \sref{sec:siteexpanyop}, and then contracted with $\Mop$, the matrix representation of the operator. 
To compute the unitary matrices involved, it suffices to recognise that $F$ moves and braids are unitary transformations in the space of labelled tree diagrams.
Identifying the leg on which the unitary matrix is to be inserted, 
the relevant region of the space of labelled diagrams is then enumerated by the multi-index which can be associated with this leg (compare \fref{fig:exampleedges}).

\begin{figure}
\includegraphics[width=246.0pt]{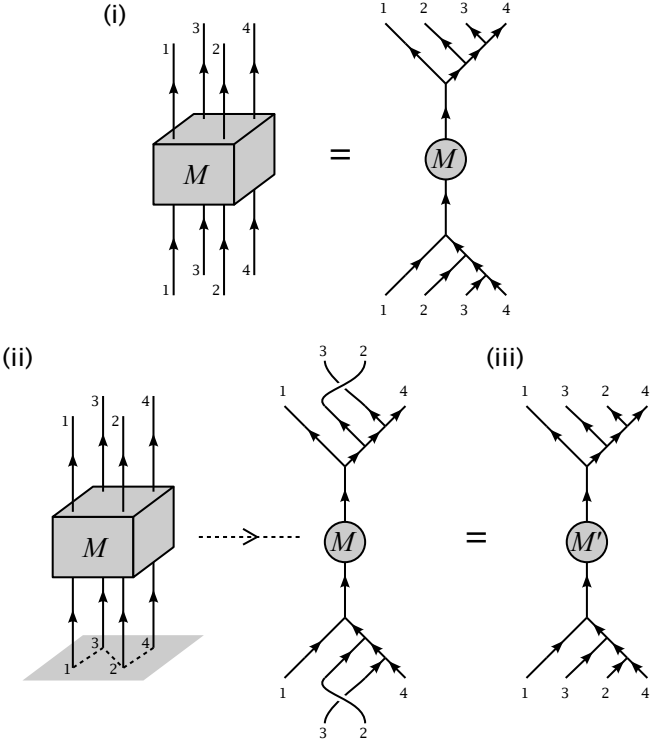}
\caption{(i) An operator $\hat M$ acting on sites on a 2-D lattice is defined with respect to some arbitrary linear ordering of these sites. (ii) When manipulating the tensor network, it may on occasion be computationally convenient for the lattice to be linearised according to some alternative linearisation scheme. In this example, the imposed linearisation scheme is indicated by the dotted line. To apply $\hat M$ to a different linearisation of the lattice may require braiding. The orientation of the braids can be determined by putting the fusion tree of (i) onto the 2-D lattice, then smoothly deforming the lattice into a chain in accordance with the linearisation prescription. (iii) The unitary matrices corresponding to the required $F$ moves and braiding operations may be absorbed into $\hat M$, defining a new operator $\hat M'$ on the linearised lattice.\label{fig:absorbbraid}}
\end{figure}

Braiding is of particular importance when working in two dimensions, as an operator will necessarily be defined with respect to some arbitrary linear ordering of its legs, and when manipulating a tensor network it may be necessary to map between this original definition and other equivalent definitions, corresponding to different leg orderings. 
For example, let $\hat M$ be a four-site anyonic operator as shown in \fref{fig:absorbbraid}(i), which we wish to apply to a 2-D lattice. For the indicated linearisation of this lattice, application of $\hat M$ will require braiding as shown in \fref{fig:absorbbraid}(ii).
By evaluating the unitary transformations corresponding to these braids and absorbing them into $\Mop$, we may define a new operator $\hat M'$ which acts directly on the linearised lattice without any intervening manipulations of the fusion/splitting trees.

We will also frequently wish to deal with tensor legs which bend vertically through 180 degrees. 
If working with an anyon model that has non-trivial Frobenius--Schur indicators, then 
indicator flags must be applied to all bends. 
Like $F$ moves and braiding, the reversal of a Frobenius--Schur indicator flag is a unitary transformation
, and once again this leads to the introduction of a unitary matrix which can be absorbed into a nearby existing tensor. 
However, we may wish to perform other operations on bends, such as absorbing them into fusion vertices or the central matrices of anyonic operators. We may also need to move a matrix $\Mop$ across a bend. We must therefore develop the description of bends in the new diagrammatic formalism.

\begin{figure}
\includegraphics[width=246.0pt]{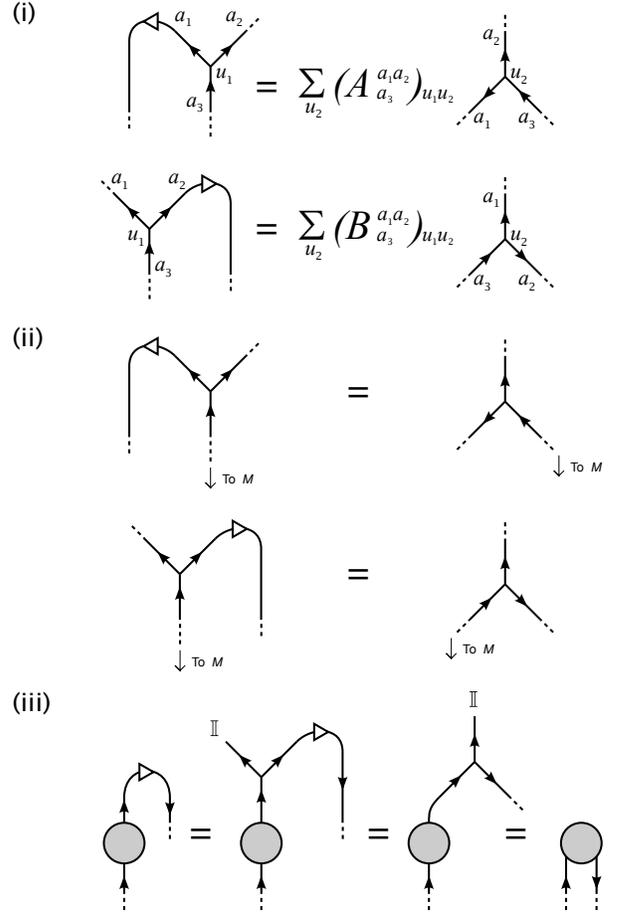}
\caption{Vertical bending of legs (i) in the standard diagrammatic notation, and (ii) in the diagrammatic notation of \protect{\sref{sec:tensordiagrammatic}}. White triangles represent Frobenius--Schur indicator flags. (iii) Legs on the matrix representations of states and operators may also absorb bends. 
\label{fig:anyonopmanip2}}
\end{figure}

In \rcite{bonderson2007} a prescription for absorbing bends into fusion vertices is given in terms of tensors $(A^{ab}_c)_{uv}$ and $(B^{ab}_c)_{uv}$, derived from the $F$ moves, and corresponding to clockwise and counter-clockwise bends respectively. 
The absorption of a clockwise or counterclockwise bend into a fusion vertex is reproduced in 
\fref{fig:anyonopmanip2}(i), and results in a vertex fusing upward- and downward-going legs.  
We now assign new tensors $(\tilde N^\mrm{CW})^{\dagger\alpha u}_{\p\dagger\gamma\beta}$ and $(\tilde N^\mrm{CCW})^{\dagger\beta u}_{\p\dagger\alpha\gamma}$ to such vertices, such that writing these transformations in the notation of \sref{sec:tensordiagrammatic} is trivial. This is shown in \fref{fig:anyonopmanip2}(ii).

Explicit expressions for the new vertex tensors $(\tilde N^\mrm{CW})^\dagger$ and $(\tilde N^\mrm{CCW})^\dagger$ may be obtained by recognising that \fref{fig:anyonopmanip2}(i) describes the action of unitary transformations 
on $\tilde N^{\dagger\alpha\beta u}_{\p\dagger\gamma}$. When the bend is counterclockwise, the corresponding unitary matrix is derived from $(A^{ab}_c)_{uv}$, and when the bend is clockwise, the unitary matrix is derived from $(B^{ab}_c)_{uv}$. We will denote these unitary matrices $A_\gamma^{\p\gamma\delta}$ and $B_\gamma^{\p\gamma\delta}$ respectively.
%
We then have
\begin{eqnarray}
(\tilde N^\mrm{CW})^{\dagger\alpha u}_{\p\dagger\gamma\beta}&=&A_\gamma^{\p\gamma\delta}\tilde N^{\dagger\alpha\epsilon u}_{\p\dagger\delta}\delta_{\epsilon\beta}\label{eq:Nbend}\\
(\tilde N^\mrm{CCW})^{\dagger\beta u}_{\p\dagger\alpha\gamma}&=&B_\gamma^{\p\gamma\delta}\tilde N^{\dagger\epsilon\beta u}_{\p\dagger\delta}\delta_{\epsilon\alpha}.\label{eq:Nbend2}
\end{eqnarray}
and conjugation describes equivalent vertices $\tilde N^{CW}$ and $\tilde N^{CCW}$ when a bend is absorbed into a splitting tree. 

\begin{figure}
\includegraphics[width=246.0pt]{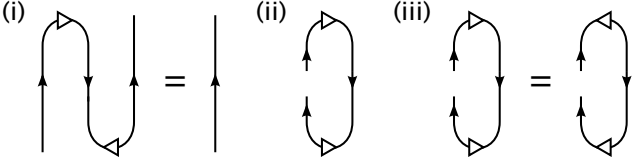}
\caption{Opposing pairs of Frobenius--Schur indicators (i) on a pair of bends equivalent to the identity, and (ii) on a pair of bends such as might be used when computing a quantum trace. (iii) As a
n anyon model can always be specified such that the Frobenius--Schur indicators are $\pm 1$, reversing a pair of contiguous opposed Frobenius--Schur indicator flags is always free
.
\label{fig:frobeniusschur}}
\end{figure}

\begin{figure}
\includegraphics[width=246.0pt]{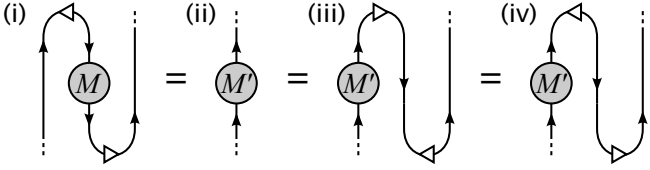}
\caption{Moving a matrix across a bend in a tensor network diagram. (i) Initial diagram. (ii) Bends are absorbed into the matrix. (iii) New bends are introduced, in accordance with \protect{\fref{fig:frobeniusschur}(i)}. (iv) A pair of contiguous, opposed Frobenius--Schur indicators are reversed, as per \protect{\fref{fig:frobeniusschur}(iii)}. The initial and final matrices $M$ and $M'$ are related as specified in Eq.~\protect{\eref{eq:movepastbend}}.\label{fig:movepastbend}}
\end{figure}

\begin{figure}
\includegraphics[width=246.0pt]{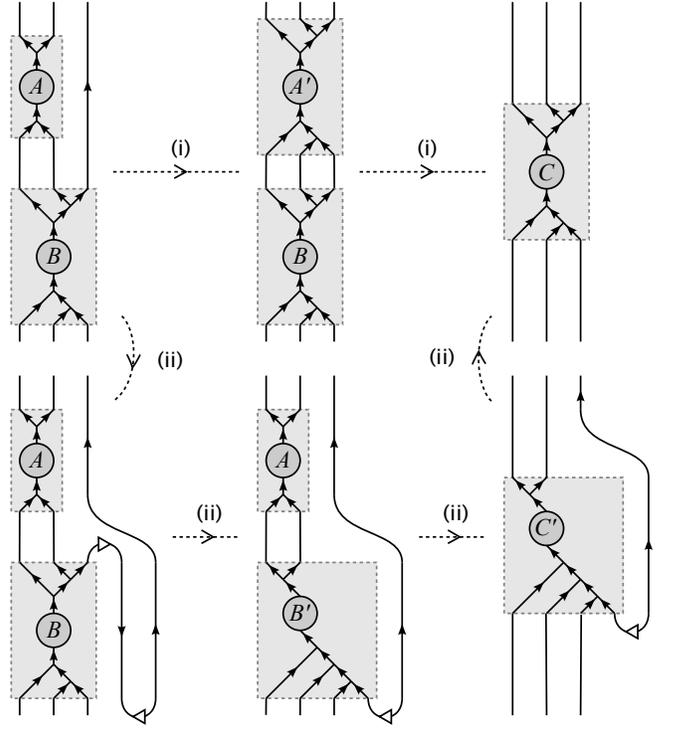}
\caption{The use of bends may permit the more efficient contraction of pairs of anyonic operators. In the sequence of events marked (i), operator $\hat A$ is first raised to the space of three sites then contracted with $\hat B$. In sequence (ii) the operators are instead contracted using bends. For many anyon models the latter approach offers a significant computational advantage.\label{fig:controps_bends}}
\end{figure}

Knowing how the absorption of bends acts on a vertex tensor, we may readily infer how the same process acts on the matrix representation of an operator. In \fref{fig:anyonopmanip2}(iii) we see a bend absorbed into the matrix $\Mop$, resulting in a new object with two lower multi-indices, $M'_{\alpha\beta}$. First we exploit the freedom to introduce fusion with the trivial charge (denoted $\mbb{I}$), with degeneracy 1. The corresponding $\tilde N^\dagger$ object takes only one value on its upper left multi-index, and is fully defined by $\tilde N^{\dagger\mbb{I}\beta 1}_{\p\dagger\gamma} = \delta^{\p\gamma\beta}_\gamma$. Absorbing the bend into this fusion vertex as per Eq.~\eref{eq:Nbend} yields $A_\gamma^{\p\gamma\delta}\delta_\delta^{\p\delta\epsilon}\delta_{\epsilon\beta} = A_{\gamma\beta}$, which may be then combined with $\Mop$ to give
\begin{equation}
M'_{\alpha\beta} = M_\alpha^{\p\alpha\gamma}A_{\gamma\beta}.
\end{equation}
In conjunction with the relationships given in \fref{fig:frobeniusschur}, this gives us the ability to move a matrix past a bend. 
An example of this is given in \fref{fig:movepastbend}, for which $M$ and $M'$ are related according to
\begin{equation}
M'^{\p\alpha\beta}_\alpha = A_{\alpha\gamma}M_\delta^{\p\delta\gamma}\varkappa_\epsilon^{\p\epsilon\delta}B^{\dagger\epsilon\beta}\label{eq:movepastbend}
\end{equation}
where $\varkappa_\epsilon^{\p\epsilon\delta}$ represents reversal of the Frobenius--Schur indicator flag on the lower bend.
Finally, bending may also allow more efficient contraction of pairs of anyonic operators, as shown in \fref{fig:controps_bends}.

\begin{figure}
\includegraphics[width=246.0pt]{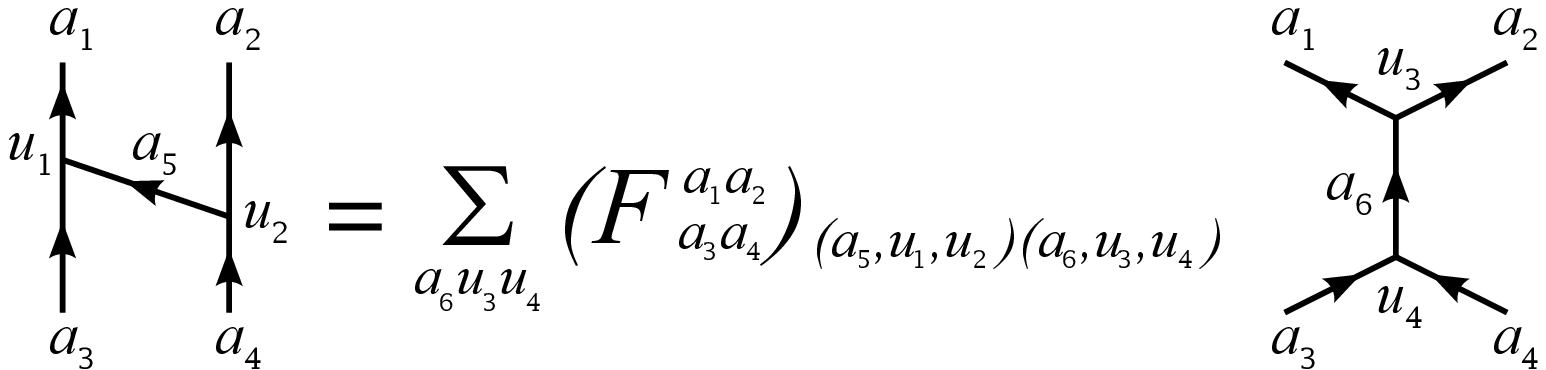}
\caption{Now that we may bend legs up and down it is customary to introduce a further type of $F$ move, derived by applying bends to the one presented in \protect{\fref{fig:anyons}}(iii). 
\label{fig:newfmove}}
\end{figure}

Having described the action of bends, it is customary also to introduce a second type of $F$ move which is described by the tensor $(F^{a_1a_2}_{a_3a_4})_{(a_5u_1u_2)(a_6u_3u_4)}$ (\fref{fig:newfmove}). 
This tensor may be derived from $(F^{a_1a_2a_3}_{a_4})_{(a_5u_1u_2)(a_6u_3u_4)}$ by bending, and as with $(F^{a_1a_2a_3}_{a_4})_{(a_5u_1u_2)(a_6u_3u_4)}$ these $F$ moves perform a transformation of the fusion tree, accompanied by the introduction of a unitary matrix which can be absorbed into the matrix representation of the operator. These unitary matrices 
correspond to the consecutive application of a bend, an $F$ move of the original type
, and a second bend whose action is the inverse of the first.

\subsection{Constructing a tensor network\label{sec:constensnet}}

\begin{figure}
\includegraphics[width=246.0pt]{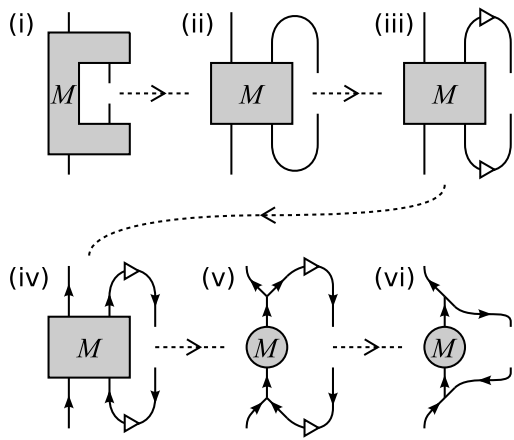}
\caption{Construction of an anyonic tensor corresponding to a normal tensor with more than three legs. (i) The original tensor. (ii) If required, any concavities are eliminated by introducing bends. (iii) Frobenius--Schur indicators are assigned to the bends. (iv) Directions are assigned to all legs, consistent with the rest of the network. 
(v) Legs are collected together into fusion and splitting trees. The central object, representing degrees of freedom of the tensor, now has less than four legs. (vi) If desired, bends can be re-absorbed into the fusion and splitting trees.
\label{fig:generalTN}}
\end{figure}

Now that we have developed a formalism for anyonic tensors,
we may convert an existing tensor network algorithm for use with anyons. 
First, the tensor network must be drawn in such a manner that every leg has a discernible vertical orientation. Although these orientations may be changed during manipulation of the tensor network, an initial assignment of upward or downward direction is required. Second, all tensors must be represented by entirely convex shapes, such as circles or regular polygons. For existing tensor network algorithms such as MERA and PEPS, this requirement is trivial. However, it is conceivable that future algorithms might involve superoperator-type objects whose graphical representations interleave upward- and downward-pointing legs. Concavities on these objects may be eliminated by replacing some of their upward-pointing legs with downward-pointing legs (or vice versa), followed by a bend (\fref{fig:generalTN}(i)-(ii)). 
A similar treatment may be applied to any superoperators which arise during manipulations of the tensor network, introducing a pair of bends as in \fref{fig:frobeniusschur}(i) and then absorbing one into the matrix representation of the object.

If working with an anyon model that has non-trivial Frobenius--Schur indicators, then indicator flags must 
be applied to all bends. 
Initial choices are a matter of convenience, and it is frequently possible to assign these indicators in opposed pairs, as shown in \fref{fig:frobeniusschur}. If these paired indicators are not flipped or are only flipped in adjacent opposed pairs 
during subsequent manipulations of the tensor network, then they may 
frequently be left implicit.

Next, if there exist charges in the anyon model which are not self-dual, a direction (represented by a solid arrow) must be assigned to every multi-index. Any tensor with more than three legs (e.g. $M$ in \fref{fig:generalTN}) is then replaced by 
a trivalent tensor network
consisting of a core object, e.g. $M\alphabeta$, which contains the free parameters of the tensor, and as many copies of $\tilde N$ or $\tilde N^\dagger$ as are required to provide the correct output legs. 
These tensors $\tilde N$, $\tilde N^\dagger$ 
correspond to vertices in the fusion and splitting trees associated with $M\alphabeta$, yielding the corresponding anyonic tensor. Objects with three legs or less can be directly identified with an anyonic tensor object carrying the appropriate number of indices (i.e. three multi-indices and a vertex index $u$), though for consistency with the methods described in Sections~\ref{sec:siteexpanyop} and \ref{sec:anyonmanip} we point out that it is possible to similarly replace three-legged objects with anyonic operators consisting of a central matrix $\Mop$ and a fusion or splitting vertex, if desired.

Any bends introduced earlier may now be reabsorbed, so that some vertices now correspond to $(\tilde N^\mrm{CW})$, $(\tilde N^\mrm{CCW})$, $(\tilde N^\mrm{CW})^\dagger$, and $(\tilde N^\mrm{CCW})^\dagger$. This step, however, is optional as it may be more convenient for subsequent manipulations of the tensor network if the bends are left explicit. 
The anyonic tensors are 
then connected precisely as in the original ansatz. 


%

Manipulations of the anyonic tensor network are equivalent to those performed on the spin version of the ansatz, differing only
in that the degrees of freedom of the tensor network are now expressed entirely by the at-most-trivalent central objects, and certain topological elements such as braids and vertical bends must be accounted for in accordance with the prescriptions of \sref{sec:anyonmanip}. 
These changes may naturally imply minor changes to the manipulation algorithms, and we will see examples of this in the {1-D} MERA. Similar considerations will 
apply 
to other tensor network algorithms. 



Our construction of an anyonic tensor network draws upon two important elements which have previously been observed in other, simpler, physical systems:
\begin{enumerate}
\item Tensors in the ansatz exhibit a global symmetry, which may be nonabelian. Exploiting a nonabelian symmetry requires that the ansatz be written in the form of a trivalent tensor network. This has previously been observed and implemented for nonabelian Lie group symmetries such as $SU(2)$\citestop{singh2009}
\item Tensors in the ansatz must be able to account for non-trivial exchange statistics. This has previously been observed in the simulation of systems of fermions\citecomma{corboz2010,kraus2010,pineda2010,corboz2009,barthel2009,shi2009,pizorn2010,gu2010} where efficient implementation of particle statistics can be achieved through the use of ``swap gates''\citestop{pineda2010,corboz2009,barthel2009,pizorn2010} 
\end{enumerate}
In both cases, anyonic tensor networks extend the concepts introduced in previous work. The symmetry structure of an anyon model may be a quantum group, for example a member of the series $SU(2)_k$, $k\in\mbb{Z}^+$, rather than having to be a Lie group, 
and this permits representation of nonabelian anyonic systems whose Hilbert space does not admit decomposition into a tensor product of local Hilbert spaces. Similarly, anyonic braiding may be implemented using a generalisation of the fermionic ``swap gate'' formalism. When braiding, particle exchange may introduce transformation by a unitary matrix rather than by a sign, and efficient implementation of the resulting swap gates is particularly important for the simulation of 2-D systems. 

Although anyonic systems pose a number of unique challenges, we see that these are addressed by developments based on existing techniques, and we therefore anticipate that the resulting generalisations of existing tensor network ans\"atze 
should still be capable of accurately representing the states of an anyonic system.




\subsection{Contraction of anyonic tensor networks}

The techniques described in Secs.~\ref{sec:siteexpanyop} and \ref{sec:anyonmanip} ($F$ moves, braids, bending of legs, elimination of loops, diagrammatic isotopy, flipping of Frobenius--Schur indicator flags, and the use of $\tilde N^{(\dagger)}$ tensors) suffice to contract any network of anyonic tensors written in the form of matrices with degeneracy indices, and unlabelled trees. Through careful application of these techniques, and avoiding at all times processes which would yield a tensor with more than three legs, the matrix representations of any pair of contiguous tensors in a network may always be brought into conjunction such that their multi-indices can be contracted in the manner of Eq.~\eref{eq:contractmatrices}, and any tensor network may be contracted by means of a sequence of such pairwise contractions.

That a tensor network may represent a system of anyons in this way
is possible because throughout the anyonic tensor network, each value of a degeneracy index is associated with a specific labelling of the corresponding unlabelled tree. Consequently it is always possible 
to fully reconstruct any operation in terms of the more verbose representation of \fref{fig:operators}.

An anyonic tensor network is therefore fully specified merely by the unlabelled tree (with Frobenius--Schur indicator flags if required), and the values and locations of the matrix representations of its tensors, written in the degeneracy index form. 

\section{Example: The {1-D} MERA}

\subsection{Construction\label{sec:MERAconstr}}

\begin{figure}
\includegraphics[width=246.0pt]{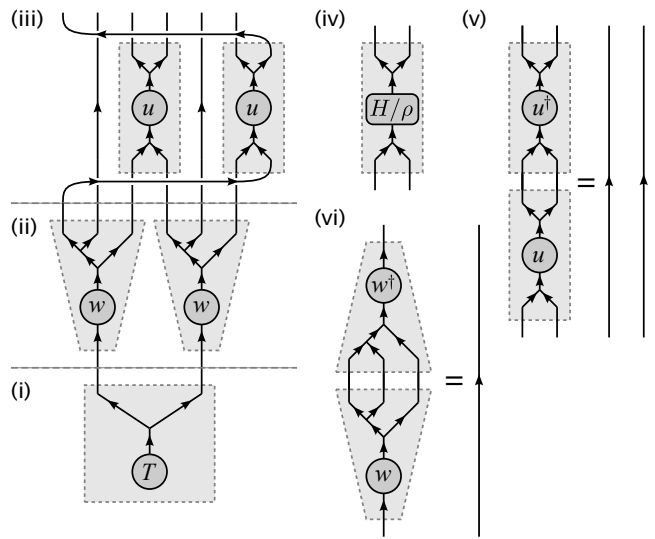}
\caption{Construction of a {1-D} ternary MERA on a periodic lattice from anyonic operators. (i) The ``top'' tensor, $\hat{T}$. (ii) Isometries, $\hat{w}$. (iii) Disentanglers, $\hat{u}$. The fusion tree representing an anyonic state (or ket) is usually drawn with the lattice sites at the top, so this MERA has been constructed ``upside down'' when compared with the diagrams in Refs.~\protect{\onlinecite{evenbly2009,vidal2010}}. This is unimportant, and we could equally well have decided to follow the convention usually adopted in tensor network algorithms, labelled the tensors in (i)-(iii) by $T^\dagger$, $w^\dagger$, and $u^\dagger$, and identified diagram (i)-(iii) as a bra.
(iv) Structure of a 2-site term in the Hamiltonian, $\hat{h}$, or a 2-site reduced density matrix, $\hat{\rho}$. 
(v) Disentanglers and (vi) isometries satisfy the relationships $\hat u\hat u^\dagger=\mbb{I}$, $\hat w\hat w^\dagger=\mbb{I}$.
\label{fig:MERAoperators}}
\end{figure}

To construct an anyonic MERA for a {1-D} lattice with $n$ sites, where $n$ satisfies $n=2\times3^k,~k\in\mbb{Z^+}$, we begin with a ``top'' tensor on a two-site lattice $\mc{L}_\tau$ whose matrix representation is of a computationally convenient size. (The top tensor is named for its position in the usual diagrammatic representation of the MERA, where diagrams with open legs at the bottom correspond to a ket. For anyons the converse convention applies, and consequently in \fref{fig:MERAoperators}(i) the ``top'' tensor is ironically located at the bottom.)

To each leg of the top tensor, we now append an isometry (\fref{fig:MERAoperators}(ii)). The matrix representations of the isometries consist of rectangular blocks, as described in \sref{sec:degenexp}, and we choose isometries whose fusion trees have three legs, so as to construct a ternary MERA\citestop{evenbly2009} Next, disentanglers are applied above the isometries.
For periodic boundary conditions this must be performed in a manner which respects the anyonic braiding rules, as shown in \fref{fig:MERAoperators}(iii). 
We identify the open legs of the resulting network as the sites of a lattice $\mc{L}_{\tau-1}$, and the rows of disentanglers and isometries may be understood as a coarse-graining transformation taking a finer-grained lattice $\mc{L}_{\tau-1}$ into a coarser-grained lattice $\mc{L}_\tau$, similar to the standard MERA. 
Note that the geometry of the periodic lattice is reflected by the connections of the disentanglers.
Specifically, whether the outside legs are braided over or under the other lattice sites reflects whether the lattice closes towards or away from the observer. 
 
The application of anyonic isometries and disentanglers is now repeated $k$ times (\fref{fig:MERAoperators}(i)-(iii) corresponds to $k=1$), until the ansatz has $n$ legs. The final row of isometries should be chosen such that each of their upper legs have the same charges and degeneracies as the sites of the physical lattice $\mc{L}_0$, and the open legs above the last row of disentanglers are identified with the physical lattice. For coarse-grained lattices $\mc{L}_1$ to $\mc{L}_\tau$, the dimensions of the lattice sites correspond to the lower legs of the isometries and are chosen for computational convenience, subject to the requirement that 
each charge sector is sufficiently large to adequately reproduce the physics of the low-energy portion of the Hilbert space. For all other legs, their charges and degeneracies are determined by requiring consistency with Eq.~\eref{eq:compounddegens}. Initial choices of which charges to represent on the ``top'' tensor and on the lower legs of the isometries, and with what degeneracies, must be guided either by prior knowledge about the physical system, or by balancing computational convenience against the inclusion of a broad and representative range of possible charges. When used in a numerical optimisation algorithm, the choice of relative weightings for the different charge sectors may often be refined by examination of the spectra of the reduced density matrices on the coarse-grained lattices, after initial optimisation of the tensor network is complete.


This concludes construction of the MERA for a state on a finite, periodic {1-D} anyonic lattice. That this tensor network does represent an anyonic state is easily seen by sequentially raising tensors, performing $F$ moves, and combining tensors, until the entire network is reduced to a single vector whose length is equal to the dimension of the physical Hilbert space, and an associated fusion tree. These then represent the state of the system as per Eq.~\eref{eq:statepsi}. The structure of this tensor network closely resembles that of the normal MERA, according to the identifications given in \fref{fig:MERAoperators}, and consequently we anticipate that it will share many of the same properties, including the ability to reproduce polynomially decaying correlators in strongly correlated physical systems.
Open lattices may also be easily represented by omitting the braided disentanglers at the edge of the diagram.

We also note that in common with the MERA for spins, the anyonic MERA may be understood as a quantum circuit, although one which carries anyonic charges in its wires. Any junction in the fusion/splitting trees may be associated with a $\tilde N$ or $\tilde N^\dagger$ tensor, and the entire network may be considered as the application of a series of gates to a Hilbert space of fixed dimension beginning mostly (or entirely, if the top tensor is considered to be the first gate) in the vacuum state, with individual gates introducing entanglement across some limited number of wires.

\subsection{Energy minimisation\label{sec:MERAoptimisation}}

The anyonic MERA can be used as a variational ansatz to compute the ground state of a local Hamiltonian. The Hamiltonian is introduced as a sum over nearest neighbour interactions, each term having the form of \fref{fig:MERAoperators}(iv), and 
optimisation of the tensor network is carried out in the usual manner\citestop{evenbly2009} Also as per usual, Hamiltonians involving larger interactions, such as next-to-nearest neighbour, can be accommodated by means of an initial exact $n$-into-one coarse-graining of the physical lattice.

\begin{figure}
\includegraphics[width=246.0pt]{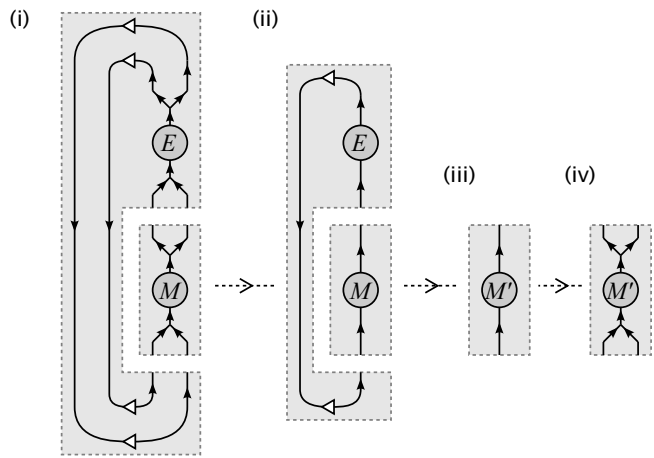}
\caption{(i) Anyonic operator $\hat E$ constitutes the environment of operator $\hat M$. 
Factors arising from the fusion and splitting trees should be evaluated and absorbed into matrices $E$ and $M$, following which (ii) matrix $E$ constitutes the environment of matrix $M$. After (iii) updating the matrix $M$ 
to $M'$, (iv) the fusion and splitting trees of $\hat M$ should be reinstated, the numerical factors associated with this process being the inverse of the fusion tree factors previously absorbed into matrix $M$. Frobenius--Schur flags in (i)-(ii) are represented by white triangles, and are not to be confused with the black arrows which indicate the orientation of lines in the fusion/splitting trees.\label{fig:environment}}
\end{figure}

As in \rcite{evenbly2009}, optimisation of the MERA then consists of repeatedly lifting the Hamiltonian from $\mc{L}_0$ to the coarse-grained lattices, updating their isometries and disentanglers, and lowering the reduced density matrix, or the top tensor and its conjugate. When lifting the Hamiltonian or lowering the reduced density matrix, then the diagrams in \rcite{evenbly2009} taken in conjunction with the key given in \fref{fig:MERAoperators} serve to describe networks of anyonic operators which, when contracted to a single operator, yield the lifted form of the Hamiltonian or lowered form of the reduced density matrix respectively. Similarly, when optimising disentanglers or isometries, the diagrams of \rcite{evenbly2009} and the identifications in \fref{fig:MERAoperators} indicate how to construct an anyonic operator which constitutes the environment of the anyonic operator being optimised. However, once the admissible ranges of charges and degeneracies on each leg have been fixed, the only optimisable content of an anyonic operator is its matrix representation. 
Consequently, the fusion and splitting tree contributions should be evaluated and absorbed into the operator and its environment,
reducing them both 
to their matrix representations, denoted $M$ and $E$ respectively (see \fref{fig:environment}).
If the singular value decomposition of $E$ is written
$E=USW^\dagger$, then the updated matrix content $M$ of the anyonic operator being optimised is given by $-WU^\dagger$, minimising the value of $\mrm{Tr}(EM)$ subject to the usual constraint for disentanglers and isometries that $\hat M\hat M^\dagger=\mbb{I}$ 
(\fref{fig:MERAoperators}(v)-(vi)). 
The fusion/splitting tree content of the operator can then be restored, along with any appropriate numerical factors that may be required.

\begin{figure}
\includegraphics[width=246.0pt]{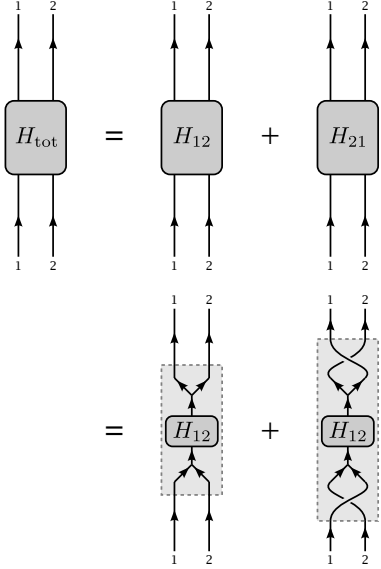}
\caption{Definition of $\hat H_{21}$ in terms of $\hat H_{12}$, on the most coarse-grained lattice ($\mc{L}_\tau$) of the translation invariant periodic MERA. Lattice $\mc{L}_\tau$ is a two-site periodic lattice.\label{fig:H21}}
\end{figure}

As with the standard MERA, the ``top'' tensor is constructed by diagonalising the total Hamiltonian on the most coarse-grained lattice, $\hat H_{\mrm{tot}}$ on $\mc{L}_\tau$. As $\mc{L}_\tau$ is a two-site lattice, the total Hamiltonian $\hat H_\mrm{tot}$ is a sum of two terms, $\hat H_{12}$ and $\hat H_{21}$. For the translation-invariant anyonic MERA, we may formally define $\hat H_{21}$ in terms of $\hat H_{12}$ as shown in \fref{fig:H21}, and the top tensor $\hat T$ (together with any factors arising from the chosen normalisation scheme) then corresponds to the lowest-energy eigenstate of $\hat H_\mrm{tot}$.

\subsection{Scale invariant MERA}

Having identified the anyonic counterparts of the tensors of the standard MERA, and described how these tensors may be lifted, lowered, and optimised, the algorithm for the scale-invariant MERA described in \rcite{pfeifer2009} may also be implemented for anyonic systems, simply by applying the dictionary of \fref{fig:MERAoperators} and the techniques described in \sref{sec:MERAoptimisation}. As with optimisation of $\hat{u}$ and $\hat{w}$, the computation of the top reduced density matrix (which is a descending eigenoperator of the scaling superoperator with eigenvalue 1) may be understood as a calculation of the matrix component $\rho_\alpha^{\p{\alpha}\beta}$ of the reduced density matrix $\hat{\rho}$. The ascending eigenoperators of the scaling superoperator, or local scaling operators of the theory, may also be computed in this manner.

\subsection{Results}

To demonstrate the effectiveness of the anyonic generalisation of the MERA, we applied it to a {1-D} critical system of anyons whose physical properties are already well known: The golden chain\citestop{feiguin2007} This model consists of a string of Fibonacci anyons subject to a local interaction. Fibonacci anyons have only two charges, $1$ (the vacuum) and $\tau$, and one non-trivial fusion rule ($\tau\times\tau\rightarrow 1+\tau$). 
The simplest local interactions for a chain of Fibonacci $\tau$ anyons are nearest neighbour interactions favouring fusion of pairs into either the $1$ channel (termed antiferromagnetic, or AFM), or the $\tau$ channel (termed ferromagnetic, or FM). Both choices correspond to critical Hamiltonians, associated with the conformal field theories $\mc{M}(4,3)$ and $\mc{M}(5,4)$ for AFM and FM couplings respectively. Individual lattice sites are each associated with a charge of $\tau$.

\begin{figure}
\includegraphics[width=246.0pt]{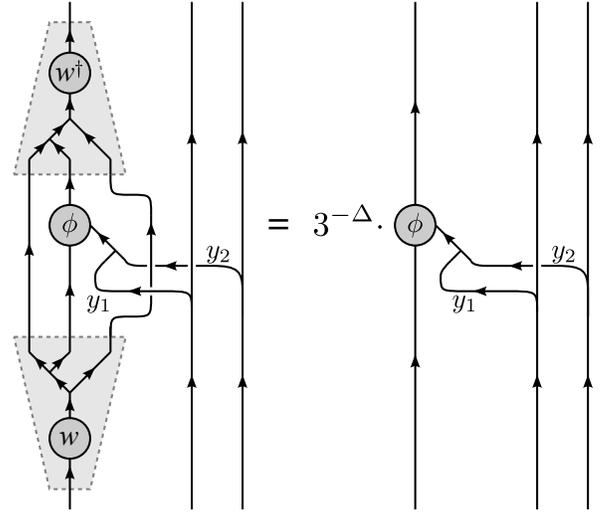}
\caption{Determination of eigenoperators ($\phi$) and associated scaling dimensions ($\Delta$) for the one-site scaling superoperator of the anyonic {1-D} MERA. Eigenoperators may be classified according to the charges on edges $y_1$ and $y_2$. One interpretation of these labels is 
that, in
addition to the sites of the {1-D} lattice, there may exist free charges lying in front of and behind the anyon chain. The labels $y_1$ and $y_2$ then represent the transfer of charge between these regions and the {1-D} lattice. 
\label{fig:scalingsuperop}}
\end{figure}

\begin{figure}
\includegraphics[width=246.0pt]{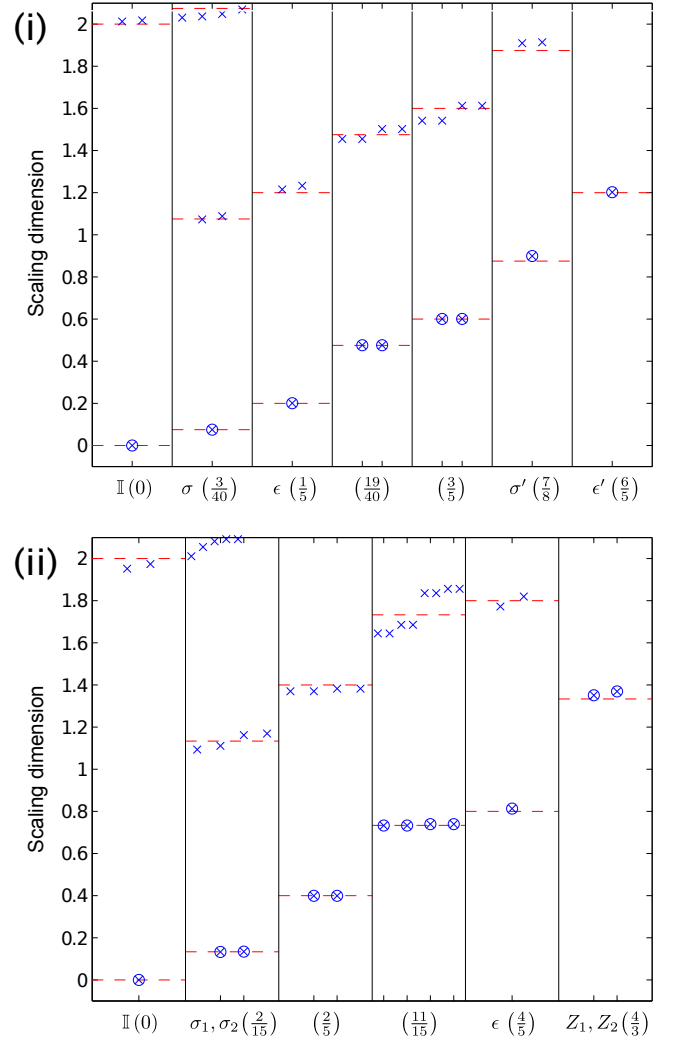}
\caption{COLOR ONLINE. Scaling dimensions of leading primary operators and their descendants, computed for 
(i) antiferromagnetic and (ii) ferromagnetic local Hamiltonians \protect{\eref{eq:Hamiltonians}} on the golden chain. 
Results are grouped into conformal towers, with a slight horizontal spread 
introduced to show the degeneracies of the descendant fields.
A circled cross indicates a primary field, and a plain cross indicates a descendant. Dashed lines indicate values predicted from CFT. 
\label{fig:results}}
\end{figure}

The AFM and FM Hamiltonians act on pairs of adjacent Fibonacci anyons. On a pair of lattice sites each carrying a charge of $\tau$, the matrix representations of the AFM and FM Hamiltonians are written
\begin{equation}
(H\alphabeta)_\mrm{AFM}=\left(\begin{array}{cc}\!\!-1\,\,&0\\\!\!\p{-}0\,\,&0\end{array}\right)
\quad
(H\alphabeta)_\mrm{FM}=\left(\begin{array}{cc}0&\p{-}0\\0&-1\end{array}\right)\label{eq:Hamiltonians}
\end{equation}
where a multi-index value of 1 corresponds to the vacuum charge, 2 corresponds to $\tau$, and the charges are non-degenerate. 
We optimised a scale-invariant MERA on the golden chain for each of these Hamiltonians
, 
and computed local scaling operators using the tensor network given in \fref{fig:scalingsuperop}. 
The operators calculated using this diagram may be classified according to the values of the charge labels $y_1$ and $y_2$, and the scaling dimensions and conformal spins which we obtained are given in Tables~\ref{tab:results} and \ref{tab:results2}, and \fref{fig:results}.
\begin{table}
\begin{tabular}{|c|c|c|}
\hline
\multicolumn{3}{|c|}{$y_1=y_2=1$}\\
\hline\hline
Exact & Numerics & Error \\
\hline
$0$ & $0$ & $0\%$\\ 
$7/8$ & $0.8995$ & $+2.80\%$\\ 
$7/8+1$ & $1.9096$ & $+1.85\%$\\
$7/8+1$ & $1.9141$ & $+2.09\%$\\
$0+2$ & $2.0124$ & $+0.62\%$\\
$0+2$ & $2.0181$ & $+0.90\%$\\
\hline
\end{tabular}
~
\begin{tabular}{|c|c|c|}
\hline
\multicolumn{3}{|c|}{$y_1=y_2=\tau$}\\
\hline\hline
Exact & Numerics & Error \\
\hline
$3/40$ & $0.0751$ & $+0.19\%$\\ 
$1/5$  & $0.2006$ & $+0.28\%$\\ 
$3/40+1$ & $1.0730$ & $-0.19\%$\\
$3/40+1$ & $1.0884$ & $+1.25\%$\\
$6/5$   & $1.2026$ & $+0.21\%$\\ 
$1/5+1$ & $1.2156$ & $+1.30\%$\\
\hline
\end{tabular}

~

~

\begin{tabular}{|c|c|c|}
\hline
\multicolumn{3}{|c|}{$y_1=1$, $y_2=\tau$}\\
\hline\hline
Exact & Numerics & Error \\
\hline
$19/40$ & $0.4757$ & $+0.14\%$\\ 
$3/5$   & $0.6009$ & $+0.15\%$\\ 
19/40+1 & $1.4549$ & $-1.37\%$\\
19/40+1 & $1.5022$ & $+1.85\%$\\
$3/5+1$ & $1.5414$ & $-3.66\%$\\
$3/5+1$ & $1.6129$ & $+0.80\%$\\
\hline
\end{tabular}
~
\begin{tabular}{|c|c|c|}
\hline
\multicolumn{3}{|c|}{$y_1=\tau$, $y_2=1$}\\
\hline\hline
Exact & Numerics & Error \\
\hline
$19/40$ & $0.4757$ & $+0.14\%$\\
$3/5$   & $0.6009$ & $+0.15\%$\\
19/40+1 & $1.4549$ & $-1.37\%$\\
19/40+1 & $1.5022$ & $+1.85\%$\\
$3/5+1$ & $1.5414$ & $-3.66\%$\\
$3/5+1$ & $1.6129$ & $+0.80\%$\\
\hline
\end{tabular}
\caption{Scaling dimensions for Fibonacci anyons with antiferromagnetic nearest neigbour interactions on an infinite chain. Numerical values were computed using an anyonic MERA with maximum degeneracies for charges $1$ and $\tau$ of 3 and 5 respectively (denoted $\chi=[3,5]$), and are grouped according to their classification by the values of $y_1$ and $y_2$ in \protect{\fref{fig:scalingsuperop}}.
\label{tab:results}}
\end{table}
\begin{table}
\begin{tabular}{|c|c|c|}
\hline
\multicolumn{3}{|c|}{$y_1=y_2=1$}\\
\hline\hline
Exact & Numerics & Error \\
\hline
$0$ & $0$ & $0\%$\\ 
$4/3$ & $1.3514$ & $+1.36\%$\\ 
$4/3$ & $1.3695$ & $+2.71\%$\\ 
$0+2$ & $1.9519$ & $-2.41\%$\\
$0+2$ & $1.9742$ & $-1.29\%$\\
$1+4/3$ & $2.2570$ & $-3.27\%$\\
\hline
\end{tabular}
~
\begin{tabular}{|c|c|c|}
\hline
\multicolumn{3}{|c|}{$y_1=y_2=\tau$}\\
\hline\hline
Exact & Numerics & Error \\
\hline
$2/15$ & $0.1329$ & $-0.35\%$\\ 
$2/15$  & $0.1339$ & $+0.44\%$\\ 
$4/5$ & $0.8134$ & $+1.67\%$\\ 
$2/15+1$ & $1.0937$ & $-3.49\%$\\
$2/15+1$   & $1.1108$ & $-1.99\%$\\
$2/15+1$ & $1.1622$ & $+2.55\%$\\
\hline
\end{tabular}

~

~

\begin{tabular}{|c|c|c|}
\hline
\multicolumn{3}{|c|}{$y_1=1$, $y_2=\tau$}\\
\hline\hline
Exact & Numerics & Error \\
\hline
$2/5$ & $0.3993$ & $-0.18\%$\\     
$11/15$   & $0.7327$ & $-0.09\%$\\ 
$11/15$ & $0.7392$ & $+0.80\%$\\
$2/5+1$ & $1.3699$ & $-2.15\%$\\
$2/5+1$ & $1.3823$ & $-1.26\%$\\
11/15+1 & $1.6450$ & $-5.10\%$\\
\hline
\end{tabular}
~
\begin{tabular}{|c|c|c|}
\hline
\multicolumn{3}{|c|}{$y_1=\tau$, $y_2=1$}\\
\hline\hline
Exact & Numerics & Error \\
\hline
$2/5$ & $0.3993$ & $-0.18\%$\\
$11/15$   & $0.7327$ & $-0.09\%$\\
$11/15$ & $0.7392$ & $+0.80\%$\\
$2/5+1$ & $1.3699$ & $-2.15\%$\\
$2/5+1$ & $1.3823$ & $-1.26\%$\\
11/15+1 & $1.6450$ & $-5.10\%$\\
\hline
\end{tabular}
\caption{Scaling dimensions for Fibonacci anyons with ferromagnetic nearest neigbour interactions on an infinite chain. Numerical values were computed using an anyonic MERA with maximum degeneracies for charges $1$ and $\tau$ of 3 and 5 respectively (denoted $\chi=[3,5]$), and are grouped according to their classification by the values of $y_1$ and $y_2$ in \protect{\fref{fig:scalingsuperop}}. 
\label{tab:results2}}
\end{table}

Comparison of the AFM case with existing results in the literature 
show that the scaling dimensions obtained when $y_1=y_2$ 
correspond to those obtained when studying a system of anyons with a toroidal fusion diagram\citestop{feiguin2007} For a system of anyons on the torus it is possible to define an additional topological symmetry\pratext{ }\cite{feiguin2007} and classify local scaling operators according to whether or not they respect this symmetry. Operators satisfying $y_1=y_2=1$ correspond to those which respect the topological symmetry, and those satisfying $y_1=y_2=\tau$ do not. We will discuss the interpretation of the different sectors and their relationship to anyons on the torus in a forthcoming paper\citestop{pfeifer2010a}

When $y_1\not=y_2$ the scaling operators obtained are chiral, with those obtained from $y_1=1,~y_2=\tau$ and $y_1=\tau,~y_2=1$ believed to form conjugate pairs.

\section{Summary}

Numerical study of systems of interacting anyons is difficult due to their non-trivial exchange statistics. To date, study of these systems has been restricted to exact diagonalisation, Matrix Product States (MPS) for {1-D} systems, or special-case mappings to equivalent spin chains. This paper shows how any tensor network ansatz may be translated into a form applicable to systems of anyons, opening the door for the study of large systems of interacting anyons in both one and two dimensions. As an example, this paper demonstrates how the MERA may be implemented for a {1-D} anyonic system. This ansatz is particularly important as many
{1-D} systems of anyons are known which exhibit extended critical phases\pratext{ }\prbtext{.}\cite[see e.g.][]{feiguin2007,trebst2008,trebst2008a}\pratext{.} The structure of the MERA is known to be particularly well suited to reproducing long range correlations, and the scale-invariant MERA has the additional 
advantage of providing simple and direct means of computing the scaling dimensions and matrix representations of local scaling operators. 

We applied the scale invariant MERA to infinite chains of Fibonacci anyons under antiferromagnetic and ferromagnetic nearest neighbour couplings, and identified a large number of local scaling operators. Our results for the scaling dimensions are in agreement with those previously obtained by exact diagonalisation of closely related systems, and for the relevant primary fields they are within $2.8\%$ of the theoretical values obtained from conformal field theory. 
We thus demonstrate that an anyonic MERA with $\chi=[3,5]$ permits conclusive identification of the relevant conformal field theory, and gives a level of accuracy comparable to that of the scale invariant MERA on a spin chain\citestop{pfeifer2009}

The anyonic generalisation of the {1-D} MERA presented here is useful in its own right, but the greatest significance of the approach described is that it is equally applicable to 2-D tensor network ans\"atze, and hence opens the door to studying the collective behaviour of large systems of anyons in two dimensions by numerical means, in situations where analytical solutions may not be possible.

The authors acknowledge the support of the Australian Research Council (FF0668731, DP0878830, DP1092513, APA). This research was supported in part by the Perimeter Institute for Theoretical Physics.

\emph{Note---}While preparing this paper for publication, we became aware of related work by \citeauthor{konig2010}\pratext{ }\prbtext{.}\cite{konig2010}\pratext{.} They also present the anyonic {1-D} MERA, providing proof of principle by
computing ground state energies for finite systems of Fibonacci anyons with $\chi=[1,1]$ ($s=2$ in their notation). 

%
%

\bibliography{AnyonMERA}

\end{document}